\documentclass[journal=aesccq,manuscript=article]{achemso}

\usepackage[version=4]{mhchem} 
\usepackage{graphicx}
\usepackage{ulem}
\usepackage{units}
\usepackage{amsmath,amssymb}
\usepackage[utf8]{inputenc}
\usepackage[T1]{fontenc}
\usepackage{xcolor}
\usepackage{appendix}
\usepackage{booktabs}
\usepackage{multirow}
\DeclareUnicodeCharacter{2212}{-}
\usepackage{graphicx}
\usepackage{txfonts}

\author{Julia C. Santos}
\affiliation[Laboratory for Astrophysics] {Laboratory for Astrophysics, Leiden Observatory, Leiden University, PO Box 9513, 2300 RA Leiden, The Netherlands}
\email{santos@strw.leidenuniv.nl}
\author{Joan Enrique-Romero}
\affiliation[Leiden Institute of Chemistry]{Leiden Institute of Chemistry, Gorlaeus Laboratories, Leiden University, PO Box 9502, 2300 RA Leiden, The Netherlands}
\author{Thanja Lamberts}
\affiliation[Leiden Institute of Chemistry] {Leiden Institute of Chemistry, Gorlaeus Laboratories, Leiden University, PO Box 9502, 2300 RA Leiden, The Netherlands}
\alsoaffiliation[Leiden Observatory] {Leiden Observatory, Leiden University, PO Box 9513, 2300 RA Leiden, The Netherlands}
\author{Harold Linnartz}
\affiliation[Leiden Observatory] {Laboratory for Astrophysics, Leiden Observatory, Leiden University, PO Box 9513, 2300 RA Leiden, The Netherlands}
\alsoaffiliation[Deceased] {Deceased, 31/12/2023}
\author{Ko-Ju Chuang}
\affiliation[Leiden Observatory] {Laboratory for Astrophysics, Leiden Observatory, Leiden University, PO Box 9513, 2300 RA Leiden, The Netherlands}

\title[]
  {Formation of \ce{S}-bearing complex organic molecules in interstellar clouds via ice reactions with \ce{C2H2}, \ce{HS}, and atomic \ce{H}}


\begin{document}

\begin{tocentry}





\includegraphics[scale=1]{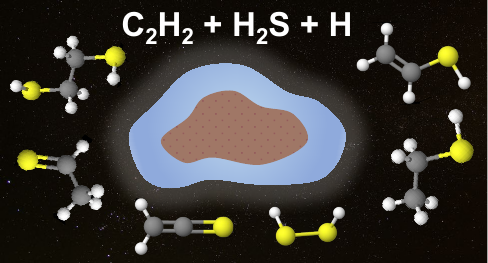}

\end{tocentry}

\begin{abstract}

The chemical network governing interstellar sulfur has been the topic of unrelenting discussion for the past decades due to the conspicuous discrepancy between its expected and observed abundances in different interstellar environments. More recently, the astronomical detections of \ce{CH3CH2SH} and \ce{CH2CS} highlighted the importance of interstellar formation routes for sulfur-bearing organic molecules with two carbon atoms. In this work, we perform a laboratory investigation of the solid-state chemistry resulting from the interaction between \ce{C2H2} molecules and \ce{SH} radicals---both thought to be present in interstellar icy mantles---at 10 K. Reflection absorption infrared spectroscopy and quadrupole mass spectrometry combined with temperature-programmed desorption experiments are employed as analytical techniques. We confirm that \ce{SH} radicals can kick-start a sulfur reaction network under interstellar cloud conditions and identify at least six sulfurated products: \ce{CH3CH2SH}, \ce{CH2CHSH}, \ce{HSCH2CH2SH}, \ce{H2S2}, and tentatively \ce{CH3CHS} and \ce{CH2CS}. Complementarily, we utilize computational calculations to pinpoint the reaction routes that play a role in the chemical network behind our experimental results. The main sulfur-bearing organic molecule formed under our experimental conditions is \ce{CH3CH2SH} and its formation yield increases with the ratios of \ce{H} to other reactants. It serves as a sink to the sulfur budget within the network, being formed at the expense of the other unsaturated products. The astrophysical implications of the chemical network proposed here are discussed. 

\end{abstract}

\section{Introduction}
\label{sec:intro}

Over 300 molecules have been detected in the interstellar and circumstellar medium to date \cite{Muller2005}, with identifications increasing at a remarkably accelerating rate within the last decade \cite{mcguire2022}. Among such detections, the so-called ``complex organic molecules'' (COMs, i.e., organic molecules with six or more atoms) are observed in various sources at different stages of star and planet formation---from prestellar cores to comets \cite{Herbst2009, Oberg2010, Bacmann2012, Biver2014, Oberg2015, Walsh2016, Herbst2017, Altwegg2017, Favre2018, Brunken2022}. Their formation routes have been extensively investigated in both gas and solid phases (see the reviews by \citet{Linnartz2015, Oberg2016, Herbst2017}). Radical-induced reactions can lead to the formation of a wide range of complex organic molecules even at temperatures as low as 10 K \cite{Watanabe2002, Linnartz2015, Butscher2015, Chuang2016, Fedoseev2017, Chuang2020, Qasim2020a, Ioppolo2021, Perrero2022}. Alternatively, interactions with photons, electrons, or cosmic rays can also trigger chemical reactions in the ice. Investigating the different pathways to forming interstellar molecules, and especially COMs, is therefore crucial to understanding the evolution of the chemical inventory of different sources.

Sulfur-bearing species, in particular, have been a long-standing issue in astrochemistry. In dense environments such as cold clouds, the (observable) gas-phase sulfur is severely depleted by up to two orders of magnitude compared to cosmic values \cite{Tieftrunk1994, Anderson2013, Vastel2018, Fuente2019, RiviereMarichalar2019}, with the bulk of its content remaining largely unknown. In addition to the sulfur depletion problem, hydrogen sulfide (\ce{H2S}) specifically also exhibits a mismatch between its predicted abundances---based on astrochemical models---and observations. It is formed very efficiently in ices via the successive hydrogenation of S atoms, producing the radical \ce{SH} as an intermediate (see, e.g., \citealt{Garrod2007, Druard2012, Esplugues2014, Vidal2017}):
\begin{equation}
    \ce{S} \xrightarrow{+ \ce{H}} \ce{SH} \xrightarrow{+ \ce{H}} \ce{H2S}.
\label{eq:H2S_form}\end{equation}
Indeed, \ce{H2S} has been detected in the gas phase towards a range of different interstellar sources \cite{Thaddeus1972, Minh1989, vanDishoeck1995, Hatchell1998, Vastel2003, Wakelam2004, Neufeld2015, LeRoy2015, Biver2015, Calmonte2016, Phuong2018}. It has also been observed in the comae of comets at the highest abundance amid all sulfur-bearing species \cite{BockeleeMorvan2000, LeRoy2015, Biver2015, Calmonte2016}. However, it has not been detected in interstellar ices so far, and estimated upper limits are of only $\leq$0.7\% with respect to water \cite{JimenezEscobar2011}. This indicates that \ce{H2S} ice must be subjected to effective destruction mechanisms, among which chemical desorption and (photo)chemical conversion seem to be particularly promising \cite{JimenezEscobar2011, JimenezEscobar2014, Oba2018, Oba2019, Shingledecker2020, Cazaux2022, Santos2023}.

A prominent loss channel involves the interaction of \ce{H} atoms with solid \ce{H2S}, resulting in the abstraction reaction:
\begin{equation}
    \ce{H2S} + \ce{H} \to \ce{SH} + \ce{H2}
\label{eq:HS_form}\end{equation}
which involves a barrier of $\sim1500$ K that can be overcome by quantum tunneling \cite{Lamberts2017}. This reaction has been shown to enrich the ice mantles with \ce{SH} radicals, which in turn can kick-start sulfur-bearing ice chemistry at temperatures of relevance to molecular clouds \cite{Santos2023}. On a similar context, \citet{Laas2019} found that a significant portion of the missing sulfur reservoir would consist of simple organosulfur compounds (e.g., \ce{OCS}, \ce{H2CS}, \ce{CS2}) locked-up in the ices. Solid-state reactions leading to \ce{S}-bearing molecules could thus play a significant role in unraveling the fate of interstellar sulfur.

Another important species to interstellar chemistry is the simplest alkyne acetylene (\ce{C2H2}). It has been observed with infrared instruments in the gas phase towards young stellar objects and envelopes of C-rich stars, as well as at fairly high abundances in the inner parts of protoplanetary disks and in comets \cite{Ridgway1976, Lacy1989, Lahuis2000, Gibb2007, Mumma2011, Bast2013, Pontoppidan2014, Tabone2023}. It is also suggested to be present in pre-stellar clouds \cite{Taniguchi2017, Taniguchi2019}. However, definitive detections in such environments require observations of gaseous molecules in submillimeter wavelengths, which are hindered for \ce{C2H2} due to its lack of a dipole moment. Since gas-phase formation routes cannot explain its observed abundances, sublimation from dust grains is usually suggested as a major source of gaseous \ce{C2H2} \cite{Charnley1992, Lahuis2000, Nguyen2002}. However, its main ice features at $\sim$3 $\mu$m and $\sim$13 $\mu$m overlap with \ce{H2O} and silicate bands, making its detection quite challenging \cite{Boudin1998, Knez2012}. This is particularly true for water-rich ices, as evinced by its high upper limit of 10 \% with respect to \ce{H2O} \cite{Boudin1998}.

A top-down mechanism through the energetic processing of polycyclic aromatic hydrocarbons (PAHs) or bare carbonaceous grains has been shown experimentally to be particularly efficient in forming \ce{C2H2} (see, e.g., \citet{Jochims1994, LePage2003, West2018}). Additionally, a bottom-up formation could be feasible through the diffusion and reaction of C atoms on dust grains yielding \ce{C2}, and its subsequent hydrogenation \cite{Tsuge2023}. The widely detected \ce{C2H} radicals (e.g., \citet{Padovani2009, Sakai2013, Kastner2014}) could also contribute to forming \ce{C2H2} upon adsorption onto ice grains followed by hydrogenation. Its gas-phase abundances in prestellar cores can however be lower than \ce{C2H2} counterparts detected towards hot cores by up to one order of magnitude \cite{Lahuis2000, Padovani2009}, which might limit its role as a dominant \ce{C2H2} precursor. As the cloud becomes denser and the residence time of H atoms on the dust surfaces increases, \ce{C2H2} may be hydrogenated to form \ce{C2H4}, \ce{C2H6}, and the radicals in-between 
\cite{Hiraoka2000, Kobayashi2017, Molpeceres2022}.

The reactive nature of the triple bond in \ce{C2H2} makes it a versatile precursor to form interstellar molecules with a carbon backbone (see, e.g., the routes proposed by \citet{Charnley2004, Molpeceres2022}). Indeed, laboratory investigations of its reaction with thermalized \ce{OH} radicals and \ce{H} atoms on interstellar-ice analogues resulted in the formation of a variety of O-bearing COMs, such as acetaldehyde (\ce{CH3CHO}), vinyl alcohol (\ce{CH2CHOH}), ketene (\ce{CH2CO}), and ethanol (\ce{CH3CH2OH}) \cite{Chuang2020}. Furthermore, energetic processing of \ce{C2H2} ices mixed with other interstellar-relevant compounds (e.g., \ce{H2O}, \ce{CO}, \ce{CO2}, and \ce{NH3}) have also been explored experimentally, probing efficient formation mechanisms to a wide range of O- and N-bearing molecules \cite{Hudson2003, Bergner2019, Chuang2021, Canta2023, Zhang2023, Chuang2024}. Its reactivity with sulfur-bearing species, however, remains to be explored under interstellar cloud conditions.

In a recent computational study, \citet{Molpeceres2022} suggested a general mechanism in which \ce{C2H2} molecules could provide the carbon backbone to form a series of different organic molecules by reacting with a range of open shell species (e.g., \ce{H}, \ce{OH}, \ce{NH2}) on interstellar-ice surfaces. Among other cases, they speculate that its interaction with \ce{H} atoms followed by \ce{SH} radicals could serve as a potential route to forming S-bearing COMs in the solid state, albeit without exploring it computationally. In the present study, we conduct an experimental investigation of the formation of S-bearing COMs via the interactions of \ce{C2H2} with \ce{H} atoms and \ce{SH} radicals. We also employ quantum-chemical calculations to elucidate the reaction network behind our laboratory results. In Section \ref{sec:methods}, the experimental setup and computational methods are described. The main results are presented and discussed in Section \ref{sec:results_diss}, and their astrophysical implications are elaborated in Section \ref{sec:astro}. In Section \ref{sec:conc} we summarize our main findings.

\section{Methods}
\label{sec:methods}
\subsection{Experimental methods}
\label{subsec:exp_methods}

This work is executed with the ultrahigh vacuum (UHV) setup SURFRESIDE$^3$, which has been described in detail elsewhere \cite{Ioppolo2013, Qasim2020b}. Here, only the relevant information is presented. A gold-plated copper substrate is mounted on the tip of a closed-cycle helium cryostat at the center of the main chamber, which operates at a base pressure of $\sim5\times10^{-10}$ mbar. The substrate temperature can be varied between 8 and 450 K using resistive heaters, and is monitored by two silicon diode sensors with a relative accuracy of 0.5 K. Gases of \ce{H2S} (Linde, purity 99.5\%) and \ce{C2H2} (Linde, 5\% in Helium) are simultaneously admitted into the chamber through all-metal leak valves. Concomitantly, a hydrogen atom beam source (HABS \cite{Tschersich2000}) generates H atoms that are subsequently cooled down by colliding with a nose-shaped quartz pipe before reaching the substrate. Ice is deposited at 10 K and monitored by Fourier-transform reflection-absorption infrared spectroscopy (FT-RAIRS), with 1 cm$^{-1}$-resolution spectra acquired in the range of 700 to 4000 cm$^{-1}$. After deposition, the sample is heated at a ramping rate of 5 K min$^{-1}$ during temperature-programmed desorption experiments (TPD). The sublimated ice species are immediately ionized by electron impact with an energy of 70 eV and are recorded by a quadrupole mass spectrometer (QMS), whilst the solid phase is concurrently monitored by RAIRS.

To quantify the abundances of products, two approaches are adopted. In the case of infrared spectroscopy, the IR integrated absorbance ($\int Abs(\nu)d\nu$) of the species in the ice can be converted to absolute abundance using a modified Beer-Lambert law:

\begin{equation}
    N_X=\ln10\frac{\int Abs(\nu)d\nu}{A'(X)}
    \label{eq:N_RAIRS}
\end{equation}

\noindent where $N_X$ is the column density in molecules cm$^{-2}$ and $A'(X)$ is the apparent absorption band strength in cm molecule$^{-1}$ of a given species. For \ce{H2S}, we utilize $A'(\ce{H2S})_{\sim2553\text{ cm}^{-1}}\sim(4.7\pm0.1)\times10^{-17}$ cm molecule$^{-1}$, as measured for our reflection-mode IR settings using the laser-interference technique \cite{Santos2023}. The band strength of \ce{C2H2} is derived by multiplying the $A$ value in transmission mode reported by \citet{Hudson2014} ($A(\ce{C2H2})_{\sim3240\text{ cm}^{-1}}\sim2.39\times10^{-17}$ cm molecule$^{-1}$) by a transmission-to-reflection conversion factor of 3.2 measured with the same experimental setup (see \citet{Santos2023} for the case of \ce{H2S}, that was later combined with the value for \ce{CO} to derive the averaged conversion factor).

Due to the higher sensitivity of the QMS compared to the RAIRS employed in this work, some of the chemical species formed in our experiments were only detectable by the former technique. In order to quantify their relative abundances, column density ratios ($N_{\ce{X}}/N_{\ce{Y}}$) can be derived from the QMS data by the expression \cite{Martin-Domenech2015}: 

\begin{equation}
    \frac{N_{\ce{X}}}{N_{\ce{Y}}}=\frac{A(\text{m/z}(X))}{A(\text{m/z}(Y))}\cdot\frac{\sigma^+(\ce{Y})}{\sigma^+(\ce{X})}\cdot\frac{I_F(\text{z}(Y))}{I_F(\text{z}(X))}\cdot\frac{F_F(\text{m}(Y))}{F_F(\text{m}(X))}\cdot\frac{S(\text{m/z}(Y))}{S(\text{m/z}(X))}
    \label{eq:N_QMS}
\end{equation}

\noindent where $A$(m/z) is the integrated desorption signal of a given mass fragment, $F_F$(m/z) is its fragmentation fraction; and $S$(m/z) is the corresponding sensitivity of the QMS. Moreover, $\sigma^+$ denotes the molecule's electronic ionization cross-section; and $I_F$(z) is the fraction of ions with charge z (here corresponding to unity). The parameters employed for each of the products found in this work are summarized in Table \ref{tab:params_QMS}. Given the overall lack of experimental values, the ionization cross sections are estimated based on the molecule's polarizability volume ($\alpha$) by the empirical correlation \cite{Hudson2006, Bull2012}:

\begin{table}[htb!]
\centering
\caption{List of parameters used to quantify the products' relative abundances with the QMS.}
\label{tab:params_QMS} 
\begin{tabular}{lccc}  
\toprule\midrule
Species         &   $\alpha$ [\AA$^3$] $^a$  &   $F_F$ (m/z) $^b$ &   $S$ (m/z) $^{b, c}$\\
\midrule
\ce{CH3CHS}     &   7.617 $^d$          &   0.598 $^g$       &   0.1\\
\ce{CH2CHSH}    &   7.578 $^e$          &   0.200 $^h$       &   0.1\\
\ce{CH3CH2SH}   &   7.38 $^f$           &   0.200 $^i$       &   0.09\\
\ce{H2S2}       &   6.828 $^d$          &   0.330 $^g$       &   0.08\\
\ce{HSCH2CH2SH} &   10.503 $^d$         &   0.164 $^i$       &   0.1\\
\midrule\bottomrule
\multicolumn{4}{l}{\footnotesize{$^a$ CCCBDB}}\\
\multicolumn{4}{l}{\footnotesize{$^b$ Values are given for the molecular ions}}\\
\multicolumn{4}{l}{\footnotesize{$^c$ \citet{Chuang2018}}}\\
\multicolumn{4}{l}{\footnotesize{$^d$ Derived by group additivity methods}}\\
\multicolumn{4}{l}{\footnotesize{$^e$ Computed with the B97D3/daug-cc-pVTZ level of theory}}\\
\multicolumn{4}{l}{\footnotesize{$^f$ \citet{Gussoni1998}}}\\
\multicolumn{4}{l}{\footnotesize{$^g$ Derived in this work}}\\
\multicolumn{4}{l}{\footnotesize{$^h$ Assuming the same as \ce{CH3CH2SH}}}\\
\multicolumn{4}{l}{\footnotesize{$^i$ NIST}}\\
\end{tabular}
\end{table}

\begin{equation}
    \sigma^+_{\text{max}}(X) = c\cdot\alpha(X)
    \label{eq:ioncross}
\end{equation}

\noindent where $X$ denotes a given species and $c$ is a correlation constant of 1.48 \AA$^{-1}$. It is expected that, for organic species, $\sigma^+$($X$) at 70 eV does not vary significantly ($<5\%$) from the maximum ionization cross section ($\sigma^+_{\text{max}}$) \cite{Hudson2003b, Bull2008}. When available, fragmentation fractions are derived from NIST \footnote{https://webbook.nist.gov/chemistry/}. Otherwise they are estimated based on our QMS measurements. In the case of \ce{CH2CHSH}, isolating its peak during TPD is quite challenging (see Section \ref{subsec:VM_ET}). Thus, we assume the same fragmentation fraction as its fully-hydrogenated counterpart, \ce{CH3CH2SH}. The polarizability values are obtained from CCCBDB\footnote{NIST Computational Chemistry Comparison and Benchmark Database (CCCBDB), NIST Standard Reference Database Number 101, http://cccbdb.nist.gov/}. Given sulfur's natural isotopic abundances, the contribution from \ce{^{34}S}-bearing species to the mass signals would be $\sim$22 times weaker than the dominant \ce{^{32}S}-bearing counterparts and therefore are not further considered.

The experiments performed in this work are summarized in Table \ref{tab:exp_list}. Both molecule fluxes, as well as the \ce{H}-atom flux, have an estimated relative error of $\lesssim5\%$. As a control experiment, pure \ce{CH3CH2SH} (Sigma-Aldrich, purity 97\%) ice is grown through vapor deposition. Previous to dosing, it is purified by a series of freeze-pump-thaw cycles. The other identified products are not commercially available and therefore standard samples cannot be obtained. The instrumental uncertainties in the integrated QMS signals are derived from the corresponding integrated spectral noise for the same band width.

\begin{table*}[htb!]
\centering
\caption{Overview of the performed experiments.}
\label{tab:exp_list} 
\begin{tabular}{lcccccccc}  
\toprule\midrule
Experiment                      &   Label  &   \ce{C2H2} flux          &   \ce{H2S} flux               &   \ce{H} flux             &    \ce{C2H2}:\ce{H2S}:\ce{H}   &   Time\\
                                &          &   (cm$^{-2}$ s$^{-1}$)    &   (cm$^{-2}$ s$^{-1}$)        &   (cm$^{-2}$ s$^{-1}$)    &                                &   (min)\\                                               
\midrule
$\ce{C2H2}+\ce{H2S}+\ce{H}$     &   1      &   $5.0\times10^{11}$      &   $5.0\times10^{11}$          &   $1.0\times10^{13}$      &    1:1:20                      &   360\\
$\ce{C2H2}+\ce{H2S}+\ce{H}$     &   2      &   $7.5\times10^{11}$      &   $3.7\times10^{12}$          &   $7.5\times10^{12}$      &    1:5:10                      &   80\\
$\ce{C2H2}+\ce{H2S}+\ce{H}$     &   3      &   $8.0\times10^{11}$      &   $8.0\times10^{12}$          &   $4.0\times10^{12}$      &    1:1:5                       &   60\\
$\ce{C2H2}+\ce{H2S}+\ce{H}$     &   4      &   $6.0\times10^{11}$      &   $6.0\times10^{11}$          &   $6.0\times10^{12}$      &    1:1:10                      &   240\\
$\ce{C2H2}+\ce{H2S}+\ce{H}$     &   5      &   $2.0\times10^{11}$      &   $2.0\times10^{11}$          &   $1.0\times10^{13}$      &    0.4:0.4:2                   &   360\\
$\ce{C2H2}+\ce{H}$              &   6      &   $6.0\times10^{11}$      &                               &   $6.0\times10^{12}$      &    1:0:10                      &   240\\
\midrule
Experiment                      &   Label  &   \ce{CH3CH2SH} flux      &                               &                           &                                &   Time\\
                                &          &   (cm$^{-2}$ s$^{-1}$)    &                               &                           &                                &   (min)\\
\midrule
$\ce{CH3CH2SH}$                 &   7      &    $7.0\times10^{12}$     &                               &                           &                                &   60\\
\midrule\bottomrule
\end{tabular}
\end{table*}

\subsection{Computational methods}

We have complemented our experimental results with computational chemical simulations for selected cases. Specifically, we have employed ORCA 5.0.4\cite{neese2020orca} to run density functional theory (DFT) calculations. The density functional of choice is M062X \cite{m062xzhao2008}, combined with the def2-TZVP basis set. All the DFT calculations were done under the unrestricted formalism and the options VERYTIGHTSCF and NORI were used. The broken symmetry approximation was used in all open shell singlet systems (i.e., radical-radical reactions). We made sure M062X is an appropriate method by comparing its performance with CCSD(T)-F12/AUG-CC-PVTZ calculations in radical-molecule reactions. For these coupled cluster calculations we used Molpro\cite{molpro1,molpro2,knizia2009simplified_CCSDtF12}. The computational results can be found in Section \ref{subsec:computation_network}.

\label{subsec:comp_methods}
\section{Results and discussion}
\label{sec:results_diss}

\begin{figure}[htb!]\centering
\includegraphics[scale=0.7]{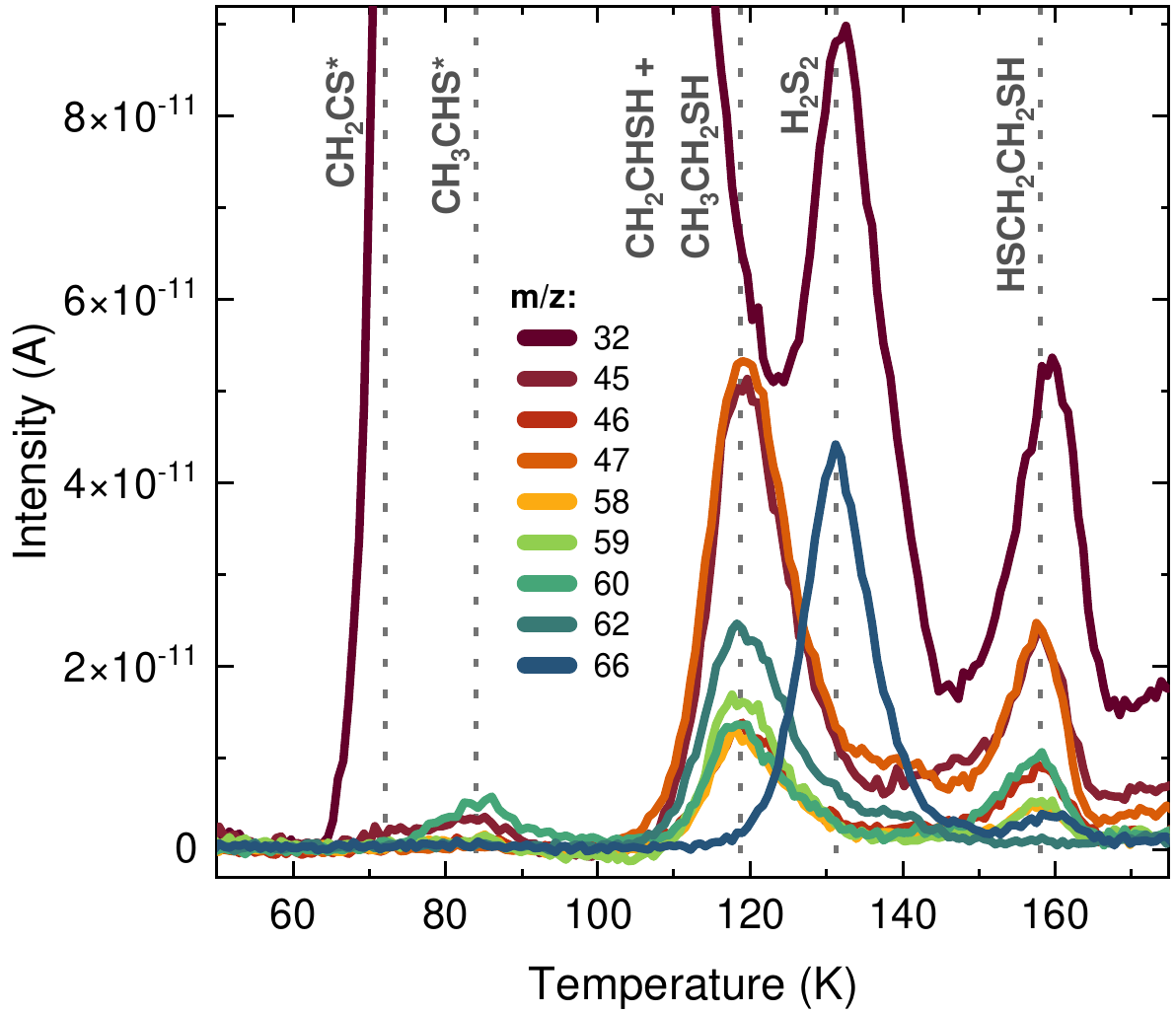}
\caption{Relevant QMS signals recorded during a TPD experiment after codepositon of \ce{C2H2}:\ce{H2S}:\ce{H} (1:1:20) at 10 K for six hours. The dotted lines mark the desorption peak of each band, and their assignments are denoted in grey. The asterisk indicates a tentative identification.}
\label{fig:TPD_all_products}
\end{figure}

\begin{figure}[htb!]\centering
\includegraphics[scale=0.5]{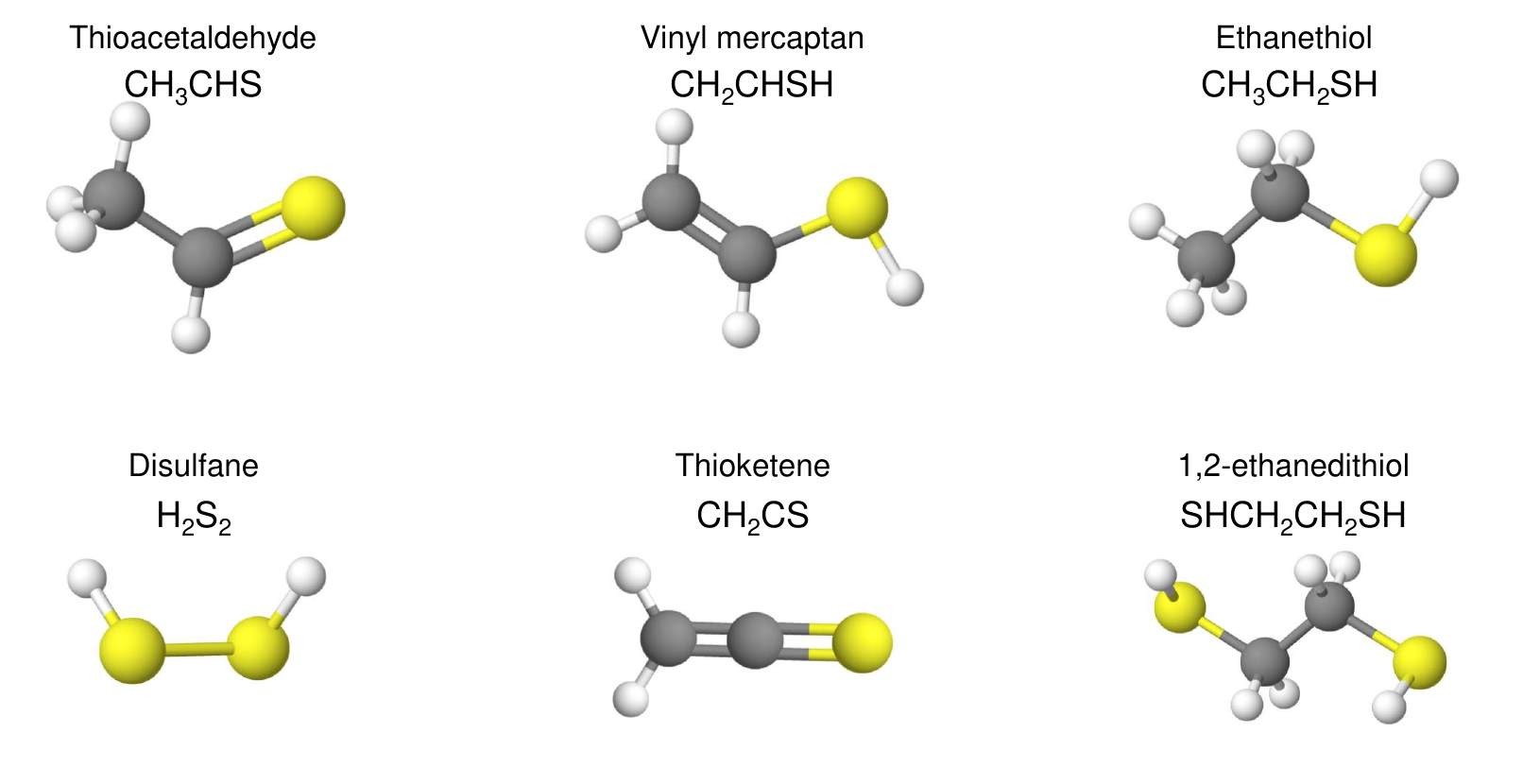}
\caption{Representative structures of the six S-bearing products identified in the experiments with \ce{C2H2} + \ce{H2S} + \ce{H}. These are provided for visualization purposes and may not depict the accurate minimum-energy geometries of the species formed in the experiments.}
\label{fig:Structures}
\end{figure}

In Figure \ref{fig:TPD_all_products}, the selected signals of m/z = 32, 45, 46, 47, 58, 59, 60, 62, and 66 recorded during TPD after the codeposition of \ce{C2H2}, \ce{H2S}, and \ce{H} (1:1:20) for six hours (experiment 1) are shown. The deposition experiment results in (at least) six reaction products as revealed by a prolific series of desorption peaks at, in decreasing order of signal intensity, $\sim119$ K, $\sim131$ K, $\sim158$ K, $\sim84$ K, and $\sim72$ K. These bands are identified as six sulfur-bearing molecules: respectively, ethanethiol (\ce{CH3CH2SH}), vinyl mercaptan (\ce{CH2CHSH}), disulfane (\ce{H2S2}), 1,2-ethanedithiol (\ce{HSCH2CH2SH}), thioacetaldehyde (\ce{CH3CHS}), and thioketene (\ce{CH2CS}). Representative structures of the identified products are shown in Figure \ref{fig:Structures} as obtained from the MolView application\footnote{https://molview.org/}. In the following subsections, the assignments of the desorption bands depicted in Figure \ref{fig:TPD_all_products} are discussed in detail.

\subsection{\ce{CH2CHSH} and \ce{CH3CH2SH}}
\label{subsec:VM_ET}

\begin{figure}[htb!]\centering
\includegraphics[scale=0.6]{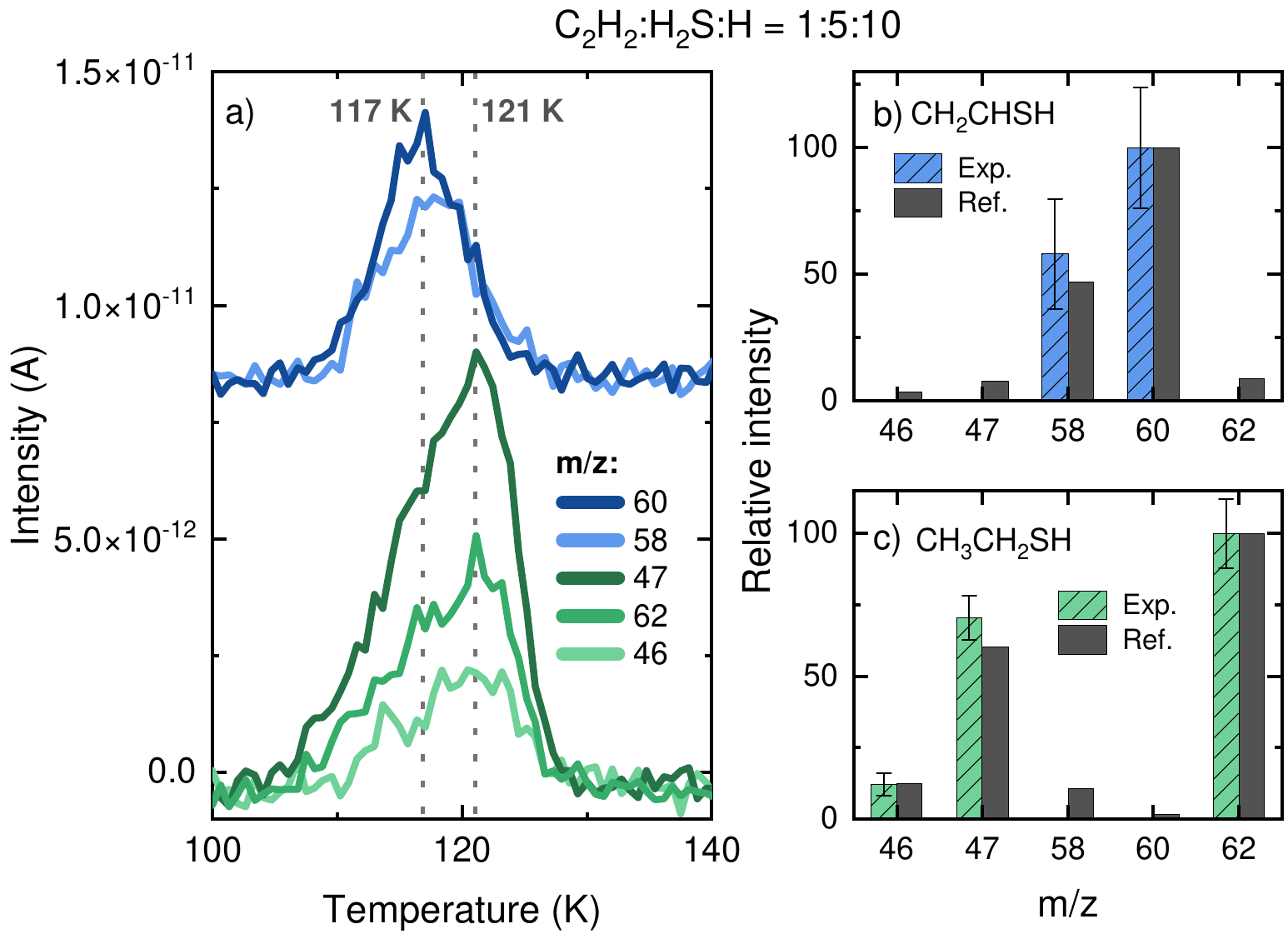}
\caption{Assignment of the TPD-QMS peaks at $\sim$117 K and $\sim$121 K as vinyl mercaptan (\ce{CH2CHSH}) and ethanethiol (\ce{CH3CH2SH}), respectively. The deposition is performed with a flux ratio of \ce{C2H2}:\ce{H2S}:\ce{H} = 1:5:10. Panel a): QMS signals showing two desorption peaks highlighted by the dotted grey lines. Panel b): Mass fragmentation pattern of the selected signals for the peak at $\sim$117 K corrected for the sensitivity of the QMS. The standard for \ce{CH2CHSH} is shown for comparison \cite{Strausz1965}. Panel c): Same as panel b), but for the peak at $\sim$121 K in comparison with \ce{CH3CH2SH} (standard measured in this work).}
\label{fig:QMS_VM_ET}
\end{figure}

The TPD of experiment 2 (\ce{C2H2}:\ce{H2S}:\ce{H} = 1:5:10) is shown in panel a) of Figure \ref{fig:QMS_VM_ET}. In contrast to experiment 1, for this mixing ratio two desorption peaks---at $\sim$117 K and $\sim$121 K---are clearly distinguishable, revealing that this band actually consists of two products. The first one is evinced by a peak in the signals for m/z = 58 and 60, whereas the second one is characterized by m/z = 46, 47, and 62. Given the elemental composition of the ice (i.e., C, S, and H atoms), these mass signals are consistent with the desorption of two sulfur-bearing organic molecules with general formula \ce{C2H_XS}. From the peak signals of m/z = 60 and 62---the largest mass-to-charge ratios detected for each peak respectively---it is inferred that the product desorbing at $\sim$117 K is described by the formula \ce{C2H4S}, and the one at $\sim$121 K is described by \ce{C2H6S}. Indeed, it is expected that the lighter species with four hydrogen atoms would be more volatile than the fully-saturated counterpart, albeit slightly. The contribution of the mass fragment m/z = 32 to these species cannot be measured since it is blended with the desorption profile of \ce{H2S} at $\sim$85 K (see, e.g., \citet{Santos2023}).

Given the proposed chemical route involving \ce{C2H2}, \ce{SH} radicals, and \ce{H} atoms, the most plausible candidates to the $\sim$117 K and $\sim$121 K peaks are, respectively, vinyl mercaptan (\ce{CH2CHSH}) and ethanethiol (\ce{CH3CH2SH}). These assignments are fortified by a comparison with the molecules' standard fragmentation pattern, although deriving the mass fragments' relative intensities is not trivial in this case. Since the full desorption bands of the two products still overlap considerably, the quantification of the contribution from each species to a given integrated mass signal is challenging. To minimize errors induced by contamination from the blended molecule, only fingerprint m/z signals uniquely associated with each species were considered in the product identification---i.e., mass fragments with a simultaneous relative intensity of $\geq$10\% in both standards of \ce{CH2CHSH} and \ce{CH3CH2SH} were excluded from the analysis, leaving only the ones with minor contributions from the adjacent feature. Panels b) and c) of Figure \ref{fig:QMS_VM_ET} show the resulting fragmentation patterns of peaks $\sim$117 K and $\sim$121 K, respectively, as well as the reference values for \ce{CH2CHSH} and \ce{CH3CH2SH}. The relative intensities of each peak match the references well, securing their assignments as \ce{CH2CHSH} and \ce{CH3CH2SH}. Furthermore, the pure \ce{CH3CH2SH} ice standard exhibits peak desorption at 119 K, in line with its identification in the \ce{C2H2} + \ce{H2S} + \ce{H} experiments. 

\begin{figure*}[htb!]\centering
\includegraphics[scale=0.63]{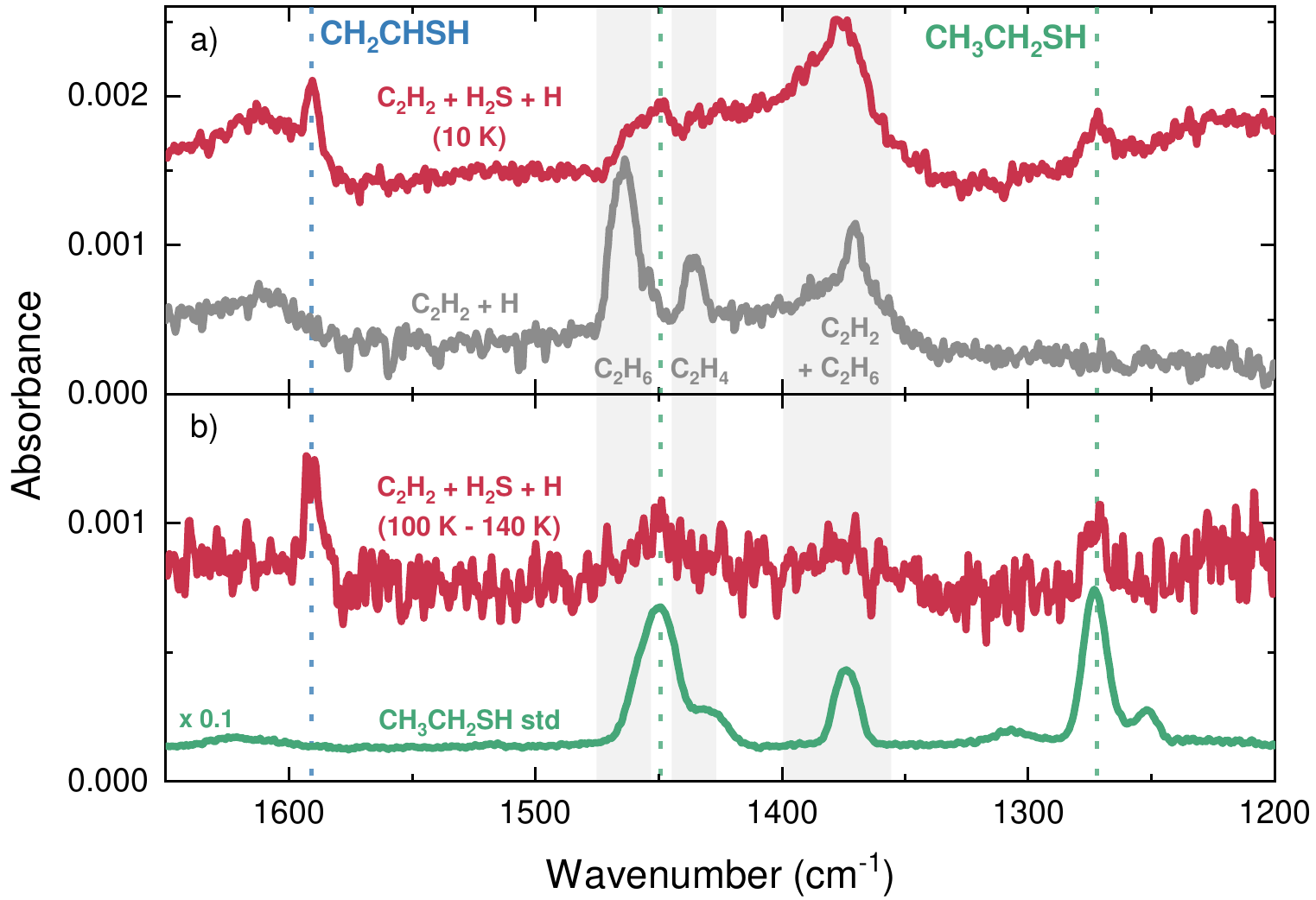}
\caption{Infrared spectra utilized to assign \ce{CH3CH2SH} and \ce{CH2CHSH}. Panel a): Spectrum recorded after codeposition of a \ce{C2H2} + \ce{H2S} + \ce{H} ice (1:1:20; red), together with a \ce{C2H2} + \ce{H} (1:10; grey) blank experiment. Panel b): difference spectrum acquired during TPD between 100 K and 140 K (red) for the \ce{C2H2} + \ce{H2S} + \ce{H} codeposition, together with a \ce{CH3CH2SH} standard spectrum (green). All depositions are performed at 10 K and the spectra are offset for clarity.}
\label{fig:IR_VM_ET}
\end{figure*}

Infrared spectroscopy is used to further assist in the identification of the most abundant products. The red IR spectrum in panel a) of Figure \ref{fig:IR_VM_ET} was acquired after a six-hours codeposition experiment with a flux ratio \ce{C2H2}:\ce{H2S}:\ce{H} = 1:1:20 (experiment 1). In the same panel, a control blank experiment of a \ce{C2H2} + \ce{H} (1:10, experiment 6) deposition is shown in grey. All depositions were performed at 10 K. In comparison with the blank experiment, the IR spectrum of \ce{C2H2} + \ce{H2S} + \ce{H} contains three additional features at $\sim1272$ cm$^{-1}$, $\sim1450$ cm$^{-1}$, and $\sim1591$ cm$^{-1}$ due to newly formed reaction products. The bands at $\sim1272$ cm$^{-1}$ and $\sim1450$ cm$^{-1}$, respectively, are consistent with the peak positions and relative intensities of the \ce{CH2} wagging and \ce{CH3} bending modes of \ce{CH3CH2SH}, as shown in panel b) by the standard spectrum in green (see also \citet{Smith1968}). The band at $\sim1591$ cm$^{-1}$ is compatible with the C=C stretching mode of \ce{CH2CHSH}, its strongest feature \cite{Almond1983}.

Infrared spectra are also obtained during the TPD experiments, which allows to correlate peaks in mass signals with the disappearance of a molecule's vibrational modes. The difference spectrum between temperatures of 100 K and 140 K of the codeposition with \ce{C2H2}:\ce{H2S}:\ce{H} = 1:1:20 (experiment 1) is shown in red in Figure \ref{fig:IR_VM_ET}, panel b). This is obtained by subtracting the spectrum measured at 140 K from the one measured at 100 K, thus highlighting the vibrational modes that disappear during this temperature range. At 100 K, \ce{C2H2} and its hydrogenation products \ce{C2H4} and \ce{C2H6} are already desorbed from the ice, and only the bands of \ce{CH2CHSH} and \ce{CH3CH2SH} are visible in the difference spectrum. These bands completely disappear from the spectra between 100 and 140 K. These desorption temperatures obtained from the IR data are consistent with the QMS measurement shown in Figure \ref{fig:QMS_VM_ET}. During warm up, the areas of the bands remain constant until reaching the temperature in which they begin to desorb, indicating that no significant change in their concentration in the ice occurs upon heating. The combination of both the infrared and the QMS data provides unambiguous evidence for the identification of \ce{CH2CHSH} and \ce{CH3CH2SH}, and signal that these molecules must be formed in the ice at 10 K and remain preserved until desorption.

\subsection{\ce{H2S2} and \ce{HSCH2CH2SH}}
\label{subsec:H2S2_DSET}

\begin{figure*}[htb!]\centering
\includegraphics[scale=0.6]{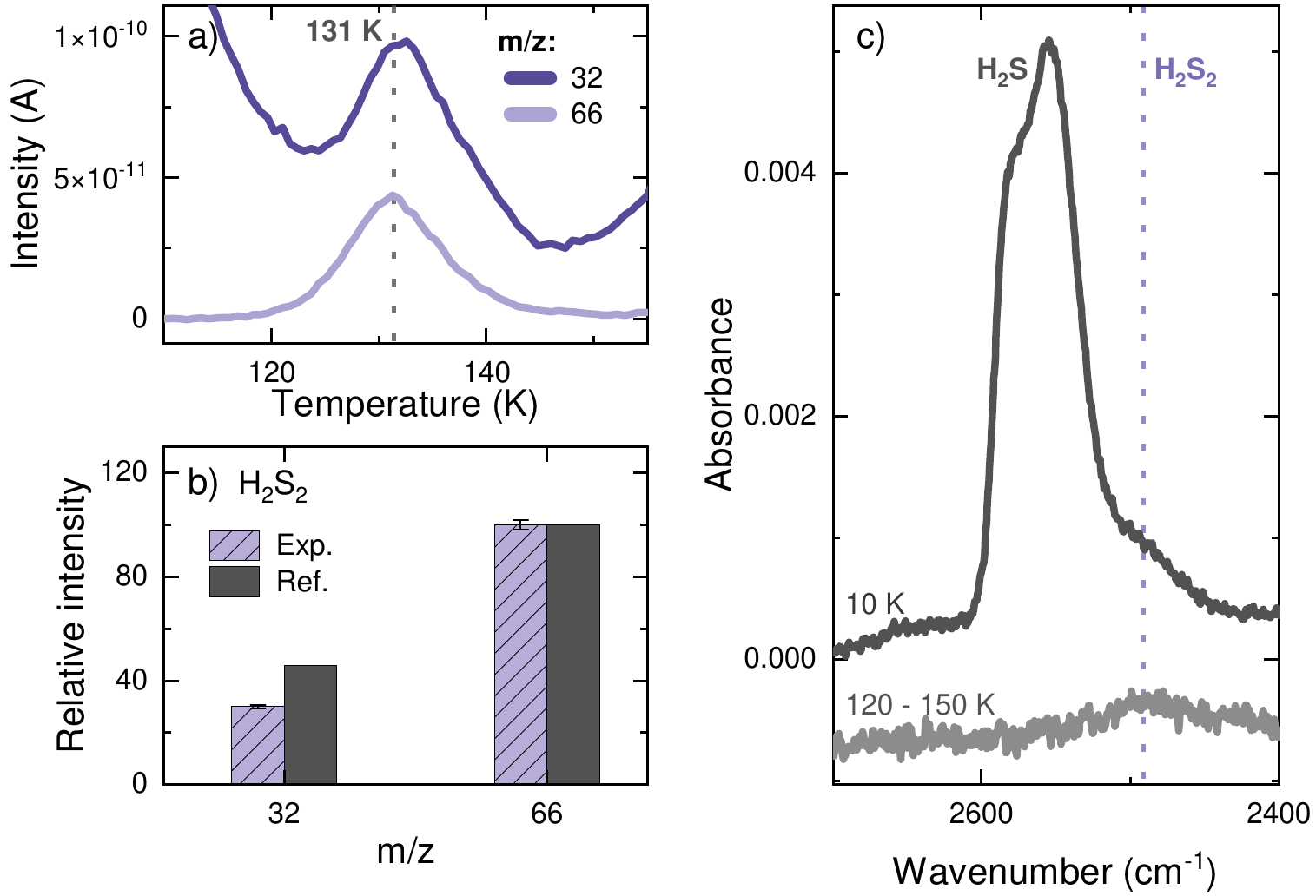}
\caption{Assignment of the TPD-QMS peak at $\sim$131 K as disulfane (\ce{H2S2}) after a codeposition experiment with a flux ratio of \ce{C2H2}:\ce{H2S}:\ce{H} = 1:1:20 (experiment 1). Panel a): QMS signals of the relevant fragments. Panel b): Mass fragmentation pattern corrected for the sensitivity of the QMS in comparison with the standard for \ce{H2S2} from \citet{Santos2023}. Panel c): Infrared spectrum at 10 K (black) and difference spectrum between 120 and 150 K (grey).}
\label{fig:H2S2}
\end{figure*}

The second strongest desorption feature in Figure \ref{fig:TPD_all_products} peaks at 131 K and contains the mass signals m/z = 32 and 66 (see Figure \ref{fig:H2S2}, panel a). Both the peak desorption temperature and the relative intensities of the fragments are in full agreement with the reference values for disulfane (\ce{H2S2}) \cite{Santos2023} (Figure \ref{fig:H2S2}, panel b). Moreover, its characteristic \ce{SH} stretching band is observed on the red wing of the \ce{H2S} feature peaking at $\sim$2491 cm$^{-1}$, as shown in panel c) of Figure \ref{fig:H2S2}. During TPD, the intensity of the \ce{H2S2} infrared feature remains constant until its complete desorption before 150 K, consistent with the desorption temperature observed for its mass fragments in the QMS data. This is highlighted by the difference spectrum between 120 K and 150 K shown in grey in Figure \ref{fig:H2S2}, panel c). Indeed, the formation of this species is unsurprising, as it has been previously shown to form at 10 K in an \ce{H2S} ice exposed to \ce{H} atoms as a result of \ce{SH} radical recombination \cite{Santos2023}.

\begin{figure}[htb!]\centering
\includegraphics[scale=0.6]{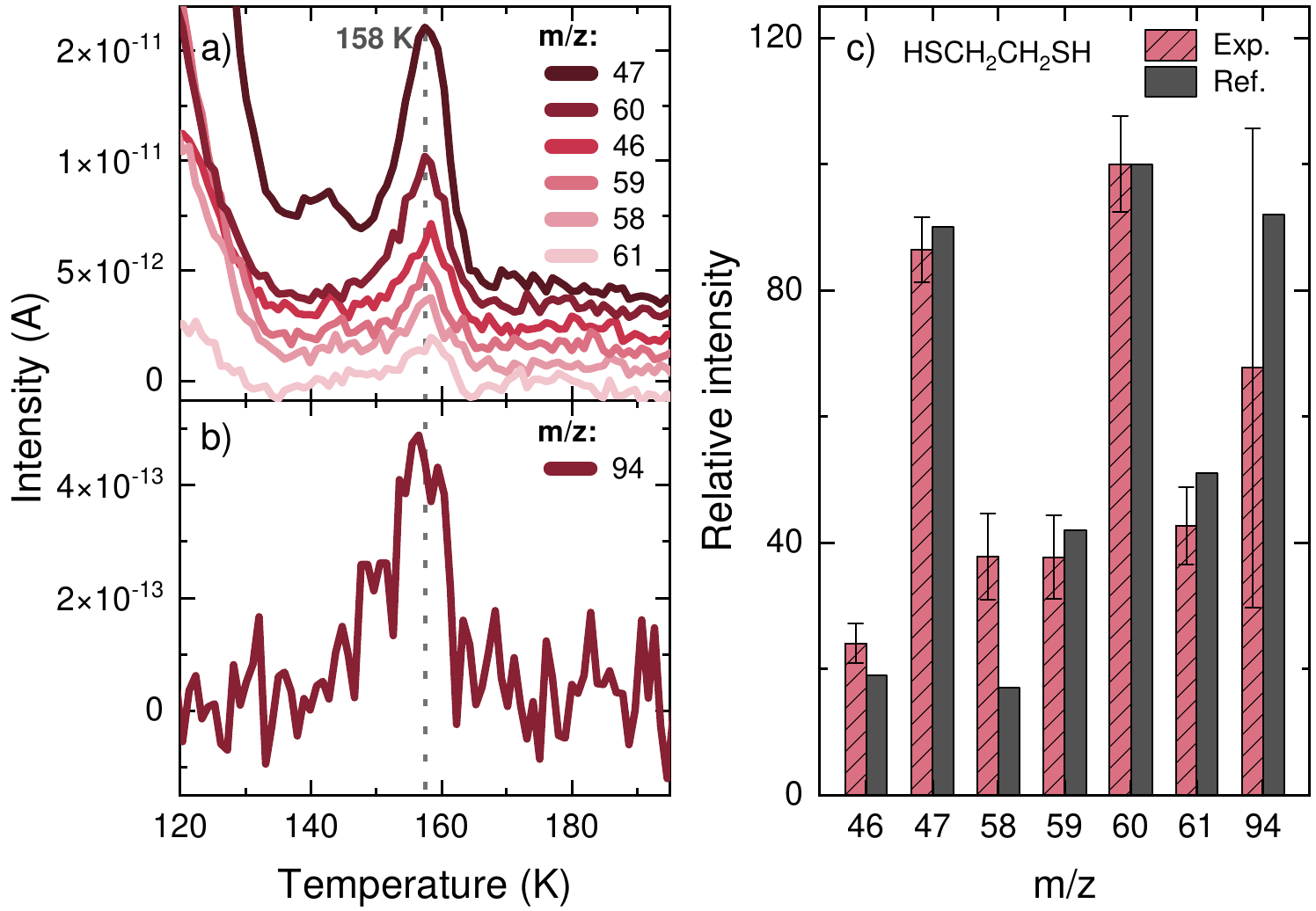}
\caption{Assignment of the TPD-QMS peak at $\sim$158 K as 1,2-ethanedithiol (\ce{HSCH2CH2SH}) after a codeposition experiment with a flux ratio of \ce{C2H2}:\ce{H2S}:\ce{H} = 1:1:20 (experiment 1). Panel a): QMS signals of the relevant fragments with m/z $\leq$ 61. Panel b): Signal for m/z = 94 measured during the same experiment. The dotted grey line indicates the peak desorption temperature. Panel c): Mass fragmentation pattern of the detected signals for the peak at $\sim$158 K corrected for the sensitivity of the QMS in comparison with the standard for \ce{HSCH2CH2SH} from NIST. The larger error bar of m/z = 94 is a consequence of the significantly lower sensitivity of the QMS at this mass-to-charge ratio.}
\label{fig:QMS_ETDT}
\end{figure}

Subsequently to \ce{H2S2}, a desorption feature appears at 158 K with contributions from m/z = 46, 47, 58, 59, 60, 61, and 94 (Figure \ref{fig:QMS_ETDT}, left panels). The mass-to-charge ratios of 92, 118, 120, and 122 were also recorded during the TPD experiment, but no increase in their signal was observed above the instrumental detection limit. Thus, it is reasonable to assume that m/z = 94 corresponds to the product's molecular ion defined by the general formula \ce{C2H6S2}. Among the possible structures associated with this formula, the most promising candidate for the assignment of the band at 158 K is 1,2-ethanedithiol (\ce{HSCH2CH2SH}). The relative intensities of the mass fragments match well with the reference values for \ce{HSCH2CH2SH} provided by NIST (Figure \ref{fig:QMS_ETDT}, right panel), and the desorption temperature is also in line with a slightly higher value of 180 K measured by \citet{Roe1998} for one layer of \ce{HSCH2CH2SH} on a molybdenum (110) surface with adsorbed carbon. 

No corresponding features of \ce{HSCH2CH2SH} are observed in the infrared spectra after deposition or during TPD, likely due to the significantly lower sensitivity of the infrared spectrometer in comparison with the QMS employed in our experiments. Moreover, this product's yield is relatively small, and its IR features heavily overlap with the other, more abundant S-bearing species sharing the same functional groups. Nonetheles, as will be discussed in detail in Section \ref{sec:astro}, the most probable route to form 1,2-ethanedithiol requires the consumption of a \ce{CH2CHSH} molecule as reactant. The fact that the infrared absorbance area of \ce{CH2CHSH} remains constant during TPD until its desorption indicates that no additional reactions involving this molecule take place as a result of the heat, thus suggesting that \ce{HSCH2CH2SH} formation should occur at 10 K.

\subsection{\ce{CH2CS} and \ce{CH3CHS}}
\label{subsec:TK_TAC}

\begin{figure}[htb!]\centering
\includegraphics[scale=0.6]{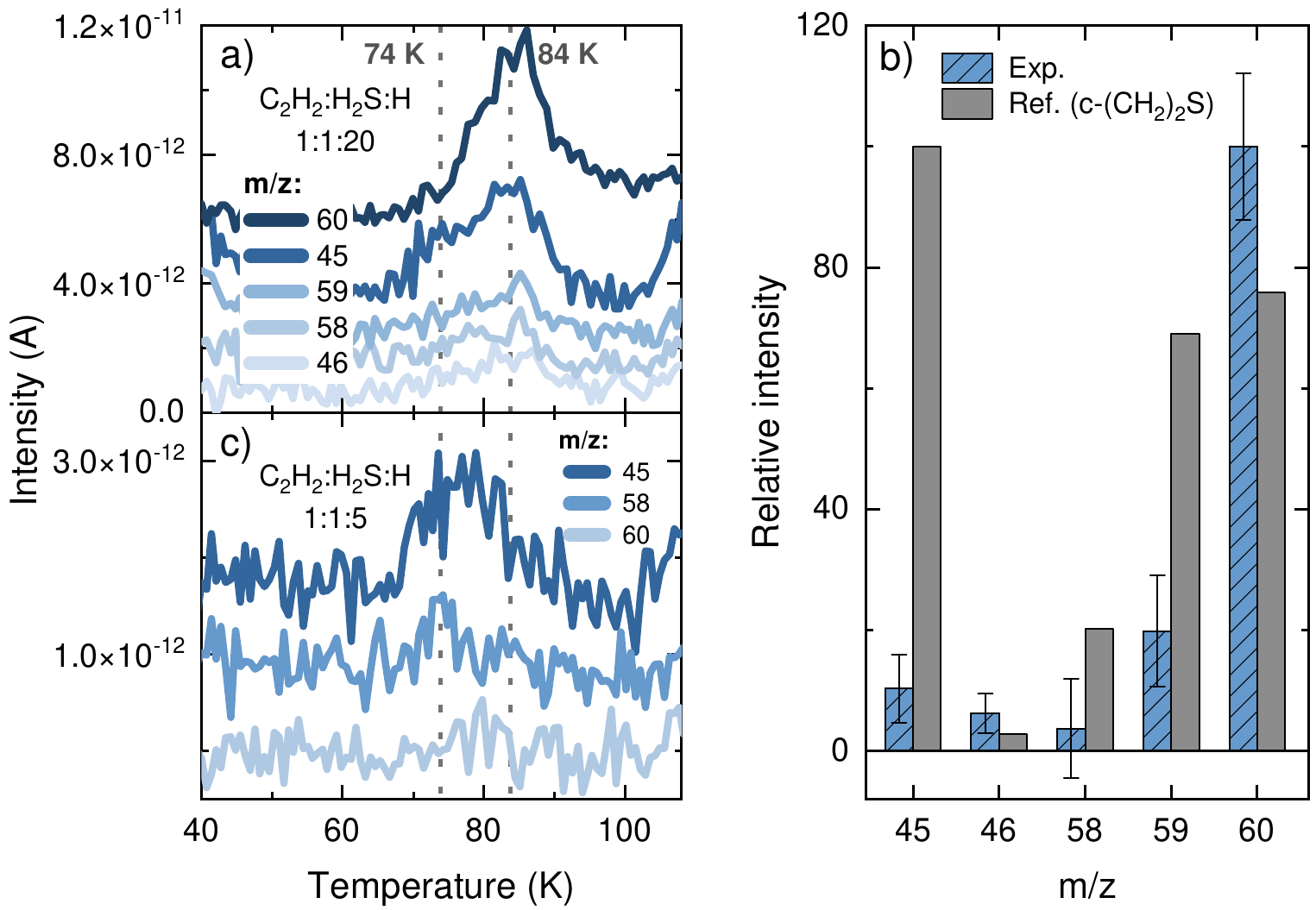}
\caption{Tentative assignments of the TPD-QMS peaks at $\sim$74 K and $\sim$84 K as, respectively, thioketene (\ce{CH2CS}) and thioacetaldehyde (\ce{CH3CHS}). Panel a): QMS signals of the relevant fragments after a codeposition experiment with a flux ratio of \ce{C2H2}:\ce{H2S}:\ce{H} = 1:1:20 (experiment 1). The dotted gray lines indicates the peak desorption temperatures. Panel b): Mass fragmentation pattern of the detected signals for the peak at $\sim$84 K corrected for the sensitivity of the QMS (blue), assigned to \ce{CH3CHS}, in comparison to the standard fragmentation pattern of \ce{c-(CH2)2S} from NIST (grey). Panel c): QMS signals of the relevant fragments after a codeposition experiment with a flux ratio of \ce{C2H2}:\ce{H2S}:\ce{H} = 1:1:5 (experiment 3), in which the \ce{CH3CHS} production is minimized. The dotted line highlights the contribution from the desorption peak at $\sim$74 K to the signals of m/z = 58 and m/z = 45.}
\label{fig:QMS_TAC_TK}
\end{figure}

Prior to the sublimation of the most abundant products, a relatively small band appears with peak desorption temperature of $\sim$ 84 K (see Figure \ref{fig:QMS_TAC_TK}, panel a). It is characterized by m/z = 45, 46, 58, 59, and 60, but without contribution from higher masses. Because its desorption temperature coincides with \ce{H2S}, we cannot probe the yield of m/z = 32 from this species. It is reasonable to assume m/z = 60 to be its molecular ion, described by the chemical formula \ce{C2H4S} (i.e., an isomer of vinyl mercaptan). Besides \ce{CH2CHSH}, three other closed-shell structures can derive from this formula: thioacetaldehyde (\ce{CH3CHS}), thiirane (\ce{c-(CH2)2S}), and thione S-methylide (\ce{CH2SCH2}). The latter is significantly less stable than the other isomers \cite{Salta2021}, and would require a cleavage of the C$\equiv$C bond in a \ce{C2H2} molecule. Thus, it is unlikely to be synthesized under our experimental conditions. Thiirane is also ruled out due to the incompatibilities of its standard fragmentation pattern from NIST and the relative intensities of the recorded mass signals (panel b) in Figure \ref{fig:QMS_TAC_TK})---in particular considering the large discrepancy in the dominant mass fragment between the two (i.e., m/z = 45 for thiirane, compared to m/z= 60 for our experiments). This leaves \ce{CH3CHS} as the most promising candidate for the desorption band at $\sim$84 K. To the best of our knowledge, no standard mass fragmentation patterns are available for this species in the literature---likely because of its extreme reactivity at room temperature. The fragments detected by the QMS are nevertheless all reasonable to arise upon electron-impact ionization of thioacetaldehyde. These are [\ce{CHS}]$^+$ (m/z = 45), [\ce{CH2S}]$^+$ (m/z = 46), [\ce{C2H2S}]$^+$ (m/z = 58), [\ce{C2H3S}]$^+$ (m/z = 59), and [\ce{C2H4S}]$^+$ (m/z = 60). Given the lack of standard measurements, this assignment is classified as tentative. Similarly to \ce{HSCH2CH2SH}, the formation of \ce{CH3CHS} also requires \ce{CH2CHSH} as a reactant (see Section \ref{subsec:computation_network} for more details), and consequently it is most likely formed at 10 K.

At $\sim$74 K, a small shoulder appears in the signal of m/z = 45 (Figure \ref{fig:QMS_TAC_TK} panel a), suggesting the desorption of another species containing the [\ce{CHS}$^{+}$] moiety. Its identification is however quite challenging given the very small yield of this product and its proximity to the more abundant \ce{CH3CHS}. For an ice mixture with mixing ratio \ce{C2H2}:\ce{H2S}:\ce{H}=1:1:5 deposited for one hour (experiment 3), the production of \ce{CH3CHS} can be significantly reduced with respect to the shoulder peak at $\sim$74 K (see Figure \ref{fig:QMS_TAC_TK} panel c), revealing an otherwise undetectable contribution from m/z = 58 to the fragmentation of the more volatile product. This mass fragment is consistent with the formula \ce{C2H2S^+} and, together with m/z = 45, suggests that the peak is due to thioketene (\ce{CH2CS}). This is however only a tentative assignment, as a more secure identifications is alas not possible given the lack of literature standards on thioketene---presumably due to its high reactivity---and the difficulty in distinguishing this product from the more abundant peak at $\sim$84 K.

\subsection{Computational results and chemical network}
\label{subsec:computation_network}

\begin{figure}[htb!]\centering
\includegraphics[scale=0.6]{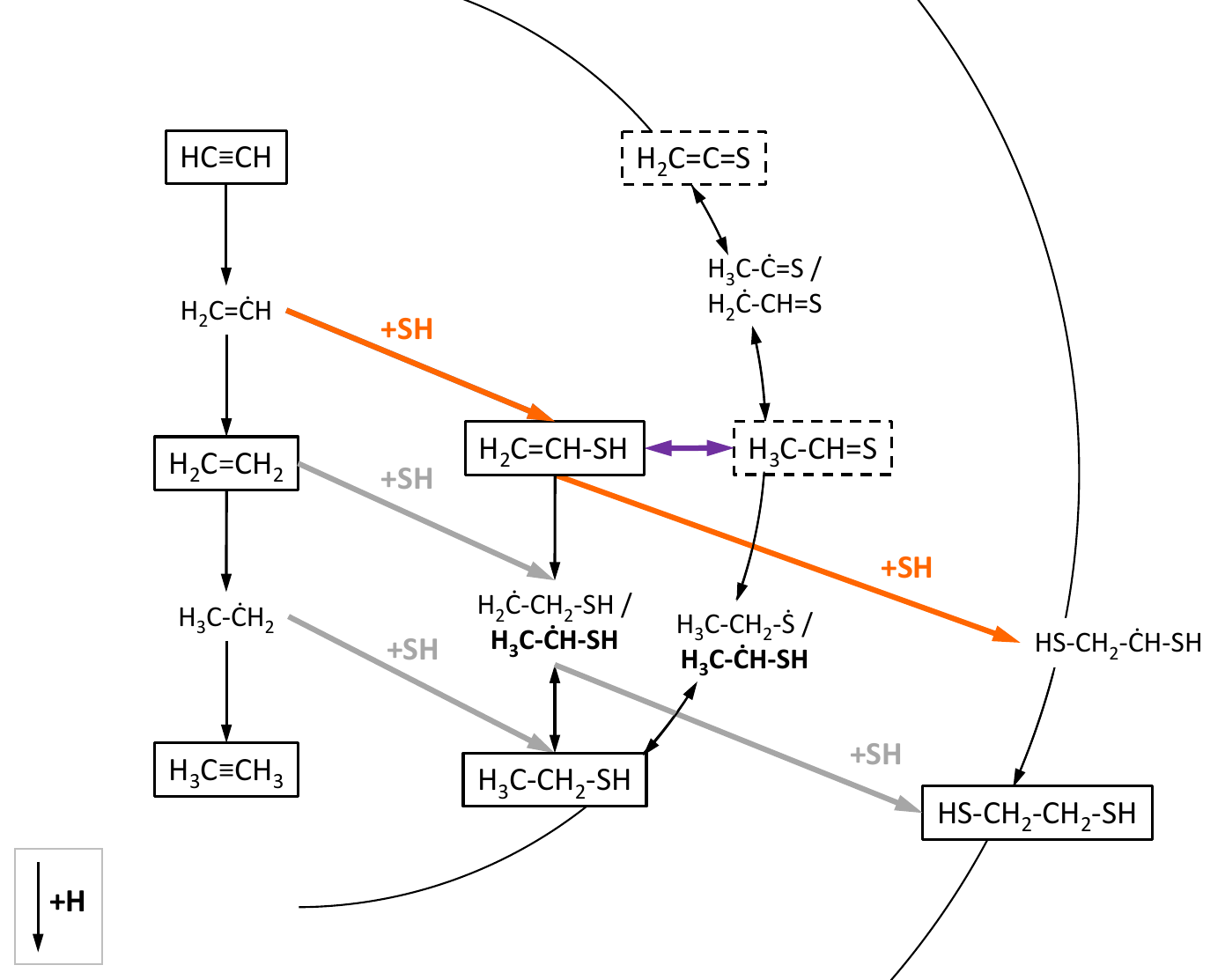}
\caption{Chemical network explored in this work. Boxes denote closed-shell species, with solid and dashed lines indicating confirmed and tentative detections, respectively. Open-shell species in bold-face are formed barrierlessly \cite{Shingledecker2022}. Reactions with \ce{SH} radicals constrained in this work are indicated by orange arrows, grey arrows display potential reactions not investigated further in this work, and black arrows represent reactions with \ce{H} atoms.The purple arrow highlights the speculated intermolecular isomerization process.}
\label{fig:scheme}
\end{figure}

We combine theoretical calculations performed in-house and from the literature to constrain the reactions at play in our experiments. In both cases, care should be taken when interpreting the activation energies since they are obtained in vacuum and therefore could differ from more representative solid-state scenarios. Nonetheless, these values are still useful for an assessment of
the feasibility of a given route. The proposed chemical network is summarized in Figure \ref{fig:scheme}. The first step in forming sulfur-bearing species under our experimental conditions involves the association of \ce{SH} to a hydrocarbon containing two \ce{C} atoms (general formula \ce{C2H_x}). The radical-molecule reaction between \ce{C2H2} and \ce{SH} is the most straightforward candidate to initiate the network:

\begin{equation}
    \ce{C2H2} + \ce{SH} \to \ce{CHCHSH}.
    \label{eq:C2H2+SH}
\end{equation}

\noindent Our DFT calculations in vacuum predict a fairly small energy barrier of only $\Delta E_a\sim$758 K for this reaction. However, due to the extremely low temperatures of our experiments (10 K) and the high mass of the \ce{SH} radical, quantum tunneling is largely restricted. Considering that Reaction \ref{eq:HS_form} is exothermic, vibrationally excited SH radicals might play a role in overcoming such a barrier. However, energy dissipation in \ce{H2O} ices takes place typically within picosecond timescales (e.g., \citet{Ferrero2023}), and we extrapolate that it will also proceed in similarly short timescales for \ce{H2S} mixed with \ce{C2H2}, since a variety of vibrational modes are present that could accept the energy that is dissipating. Thus, Reaction \ref{eq:C2H2+SH} is unlikely to initiate the chemical network probed here.

Alternatively, the network is proposed to be kick-started by the formation of \ce{C2H3} through a hydrogen addition reaction to \ce{C2H2} and its subsequent association with \ce{SH}:

\begin{equation}
    \ce{C2H2} + \ce{H} \to \ce{C2H3}
    \label{eq:C2H2+H}
\end{equation}
\begin{equation}
    \ce{C2H3} + \ce{SH} \to \ce{CH2CHSH}.
    \label{eq:C2H3+SH}
\end{equation}

\noindent Reaction \ref{eq:C2H2+H} has been previously investigated in experimental as well as theoretical works and has been shown to proceed efficiently on icy surfaces at $\sim$10 K \cite{Hiraoka2000, Miller2004, Kobayashi2017, Molpeceres2022}, with a predicted activation energy of $\sim2141-2430$ K depending on the level of theory utilized \cite{Molpeceres2022}. Despite being significantly larger than the activation energy for Reaction \ref{eq:C2H2+SH}, it does not hinder Reaction \ref{eq:C2H2+H} due to the much more efficient quantum tunneling of the light \ce{H} atoms as opposed to the heavy \ce{SH} radicals. Indeed, estimated rate constants for Reaction \ref{eq:C2H2+H} in the solid phase are of the order of $10^4$ s$^{-1}$ \cite{Molpeceres2022}. The subsequent Reaction \ref{eq:C2H3+SH} takes place between two radicals, \ce{C2H3} and \ce{SH}. The former is a non-polar species, and hence is expected to have a low binding energy to the \ce{H2S} ice. Since the ice is continuously grown, a randomized distribution of orientations is to be expected. Thus, reorientation barriers as predicted by \citet{EnriqueRomero2022} and observed by, e.g., \citet{MartinDomenech2020}, are less of a limitation. Consequently, nondiffusive interactions between \ce{C2H3} and \ce{SH} radicals in the vicinity of one another are expected to efficiently form \ce{CH2CHSH} and initiate the sulfur chemical network in Figure \ref{fig:scheme}. Such nondiffusive mechanisms have been extensively studied in the laboratory and through chemical modeling and are suggested to contribute significantly to ice chemistry under dark cloud conditions (\citet{Jin2020, Garrod2022}, and references therein).

Following its formation via Reaction \ref{eq:C2H3+SH}, \ce{CH2CHSH} can be hydrogenated to form \ce{CH3CH2SH}:
\begin{subequations}
\begin{equation}
    \ce{CH2CHSH} + \ce{H} \to \ce{CH3CHSH}
    \label{eq:CH2CHSH+H_1}
\end{equation}
\begin{equation}
    \ce{CH2CHSH} + \ce{H} \to \ce{CH2CH2SH}
    \label{eq:CH2CHSH+H_2}
\end{equation}
\end{subequations}

\begin{subequations}
\begin{equation}
    \ce{CH3CHSH} + \ce{H} \to \ce{CH3CH2SH}
    \label{eq:CH3CHSH+H}
\end{equation}
\begin{equation}
    \ce{CH2CH2SH} + \ce{H} \to \ce{CH3CH2SH},
    \label{eq:CH2CH2SH+H}
\end{equation}
\end{subequations}

\noindent which proceeds by first forming either the radical \ce{CH3CHSH} (Reaction \ref{eq:CH2CHSH+H_1}, $\Delta E_a=0$ in vacuum) or \ce{CH2CH2SH} (Reaction \ref{eq:CH2CHSH+H_2}, $\Delta E_a\sim1227$ K in vacuum) \citep{Shingledecker2022}. The subsequent hydrogen additions in Reactions \ref{eq:CH3CHSH+H} and \ref{eq:CH2CH2SH+H} proceed without an activation barrier. 

Alternatively, \ce{CH2CHSH} can also react with nearby \ce{SH} radicals to form the doubly-sulfurated radical \ce{HSCH2CHSH}:

\begin{equation}
    \ce{CH2CHSH} + \ce{SH} \to \ce{HSCH2CHSH},
    \label{eq:CH2CHSH+SH}
\end{equation}

\noindent for which no activation energy barrier was observed in our vacuum DFT calculations. The hydrogenation of this radical will lead to the formation of \ce{HSCH2CH2SH}:

\begin{equation}
    \ce{HSCH2CHSH} + \ce{H} \to \ce{HSCH2CH2SH}.
    \label{eq:HSCH2CHSH+H}
\end{equation}

\noindent Similarly, radical-radical coupling reactions between \ce{SH} +\ce{CH2CH2SH}/\ce{CH3CHSH} could also form \ce{HSCH2CH2SH} efficiently.

A third possible fate for the \ce{CH2CHSH} molecule is that it is converted to its structural isomer \ce{CH3CHS}:

\begin{equation}
    \ce{CH2CHSH} \leftrightarrow \ce{CH3CHS}.
    \label{eq:CH2CHSH_iso}
\end{equation}

\noindent Intramolecular isomerization (i.e., a transfer of a hydrogen atom within one molecule) is unlikely to occur at 10 K given its high energy barrier ($\Delta E_a \gtrsim 28000$ K in vacuum) \citep{Salta2021, Shingledecker2022}. Alternatively, transfers of hydrogen atoms in a concerted mechanism involving multiple molecules from the ice are associated with significantly lower activation energies and could arguably take place in our experiments. This process, known as ``intermolecular isomerization'', has been widely studied in both the gas and liquid phases for the oxygen-bearing counterparts \ce{CH2CHOH} and \ce{CH3CHO} \cite{Klopman1979, Capon1982, Lledos1986, DaSilva2010}, and is shown to be catalyzed by surrounding \ce{H2O} molecules \cite{Capon1982, Lledos1986}. This same mechanism has been proposed previously to explain the large abundances of \ce{CH2CHOH} and \ce{CH3CHO} in similar ice experiments involving \ce{C2H2} and \ce{OH} radicals \citep{Chuang2020}. However, \citet{Perrero2022} found this mechanism to be hampered by high activation energy barriers ($\gtrsim$ 7000 K) in pathways starting from both \ce{CH2CHOH} and its radical precursor \ce{CH2CHO}. Interestingly, theoretical calculations by \citet{Suenobu1999} predict a faster conversion of \ce{CH2CHSH} into \ce{CH3CHS} mediated by \ce{H2O} molecules in the liquid phase compared to the oxygen-bearing counterparts. They argue that this is due to the larger electric dipole moment of the transition state in the sulfur-bearing case, causing it to be more stabilized by aqueous environments. We therefore speculate that such a mechanism could also proceed in the solid state under our experimental conditions facilitated by \ce{H2S} molecules in the vicinity of \ce{CH2CHSH}, although further theoretical work is warranted to accurately constrain this possibility.

We emphasize that \ce{CH3CHS} is tentatively detected in this work and contains relatively low abundances. Thus, its suggested formation by the intermolecular isomerization mechanism is probably not very efficient. Nonetheless, assuming that it can be formed, further hydrogenation will lead to \ce{CH3CH2SH} via:

\begin{subequations}
\begin{equation}
    \ce{CH3CHS} + \ce{H} \to \ce{CH3CHSH}
    \label{eq:CH3CHS+H_1}
\end{equation}
\begin{equation}
    \ce{CH3CHS} + \ce{H} \to \ce{CH3CH2S}
    \label{eq:CH3CHS+H_2}
\end{equation}
\end{subequations}

\begin{equation}
    \ce{CH3CH2S} + \ce{H} \to \ce{CH3CH2SH}.
    \label{eq:CH3CH2S+H}
\end{equation}

\noindent Reaction \ref{eq:CH3CHS+H_1} proceeds barrierlessly, whereas Reaction \ref{eq:CH3CHS+H_2} has a very small $\Delta E_a \sim 397$ K \citep{Shingledecker2022}. Both \ce{CH3CHSH} and \ce{CH3CH2S} will be hydrogenated without an activation energy through Reactions \ref{eq:CH3CHSH+H} and \ref{eq:CH3CH2S+H} to form \ce{CH3CH2SH}. Finally, \ce{CH3CHS} can potentially undergo two hydrogen-abstraction reactions to form \ce{CH2CS}, another tentative product of this work. This alluded abstraction route warrants dedicated experimental and theoretical investigations to be confirmed, as only a tentative detections of both reactants and products are provided here.

Abstraction reactions from the radicals \ce{CH2CH2SH} and \ce{CH3CHSH} are also possible a priori, and are explored computationally here. Our DFT calculations predict that \ce{CH2CH2SH} can undergo abstraction routes induced by both \ce{H} atoms and \ce{SH} to form thiirane (\ce{c-(CH2)2S}):

\begin{equation}
    \ce{CH2CH2SH} + \ce{H} \to \ce{c-(CH2)2S} + \ce{H2}
    \label{eq:CH2CH2SH+H_2}
\end{equation}
\begin{equation}
    \ce{CH2CH2SH} + \ce{SH} \to \ce{c-(CH2)2S} + \ce{H2S}.
    \label{eq:CH2CH2SH+SH}
\end{equation}

\noindent Likewise, \ce{CH3CHSH} can also form thiirane by reacting with \ce{H}:

\begin{equation}
    \ce{CH3CHSH} + \ce{H} \to \ce{c-(CH2)2S} + \ce{H2}
    \label{eq:CH3CHSH+H_2}
\end{equation}

\noindent This structure, however, is excluded as a major product in our experiment on the basis of its mass fragmentation pattern (see section \ref{subsec:TK_TAC}). The reason behind this discrepancy between theory and experiment may be related to the former being performed in vacuum, when interactions with other species in the ice could potentially hinder the formation of \ce{c-(CH2)2S}.

Most of the hydrogenation steps in this network ultimately lead to the formation of \ce{CH3CH2SH}. This molecule could in principle be further processed by reactions with \ce{H} atoms to form the radicals \ce{CH3CH2S}, \ce{CH3CH2}, \ce{CH3CHSH}, or \ce{CH2CH2SH} \cite{Zhang2006}. The theoretical calculations by \citet{Zhang2006} predict that the hydrogen abstraction channel from the \ce{SH} group ($\ce{CH3CH2SH} + \ce{H} \to \ce{CH3CH2S} + \ce{H2}$) is the dominant route in vacuum ($\Delta E_a \sim$ 1610 K), but that the C-S bond breaking channel ($\ce{CH3CH2SH} + \ce{H} \to \ce{CH3CH2} + \ce{H2S}$) is also likely to proceed ($\Delta E_a \sim$ 1761 K). The contribution from these channels might vary when surfaces are considered (as was shown by \citet{Nguyen2023} within their work and in comparison to \citet{Lamberts2018} for the analogue reactions involving \ce{CH3SH}). We do not find any evidence (such as a peak in mass signal) for the presence of \ce{CH3CH2SSCH2CH3} (diethyl-disulfide, DEDS) in our experiments---the presumed product of the recombination of two \ce{CH3CH2S} radicals. This is rather unsurprising, since the addition routes to \ce{CH3CH2SH} proceed effectively barrierlessly and multiple reaction steps, as well as close proximity of reactants, would be needed to form DEDS. Overall, the chemical network probed here converges into forming mostly \ce{CH3CH2SH} as long as enough \ce{H} is available. This conclusion can assist in explaining astronomical observations of S-bearing COMs with two carbon atoms (or lack thereof), as will be discussed in the following Section. 

\section{Astrophysical implications}
\label{sec:astro}

Within interstellar clouds, low activation barriers are generally required for a chemical reaction to occur, as thermal hopping is not possible for most molecular radicals at the typical temperatures of those environments ($10-20$ K). The high reactivity associated with open shell species thus makes radicals important drivers of chemical complexity in such astronomical environments. This is particularly relevant for solid-state reactions facilitated by interstellar dust grains. In the case of sulfur, the \ce{SH} radical can serve as an important pivot towards building a complex sulfur-bearing inventory, especially because it is expected to be continuously formed in interstellar ices throughout a wide evolutionary span. During the earlier cloud stages, it can be readily produced by the hydrogenation of sulfur atoms accreted onto interstellar dust grains. As the density of the environment increases, most of the atomic sulfur is expected to be readily converted into \ce{H2S}. Abstraction reactions involving \ce{H} atoms and \ce{H2S} can then efficiently reform \ce{SH}, thus partially replenishing the supply of this radical in the ice.

The deuterium fractionation of \ce{H2S} observed in Class 0 sources points to an early formation in ices, before the \ce{CO} catastrophic freeze-out stage \cite{Ceccarelli2014}. Similarly, \ce{C2H2} is expected to be more abundantly present in earlier cloud stages, where atomic carbon is available to produce \ce{C2H2} via the bottom-up route, and before \ce{C2H2} is largely hydrogenated to form \ce{C2H4} and \ce{C2H6}. Reactions involving the two species and the ubiquitous hydrogen atoms are therefore feasible given that they coexist in the same ice environment. As illustrated in Figure \ref{fig:scheme}, these interactions can enrich interstellar ices with complex organic sulfur-bearing molecules. Furthermore, \ce{C2H} radicals could act as an additional source of carbon within our network upon adsorption onto ice grains. It should be noted, however, that the scheme proposed here could be further complicated in fully representative interstellar ices due to the presence of other species, in particular water.

\begin{figure}[htb!]\centering
\includegraphics[scale=0.5]{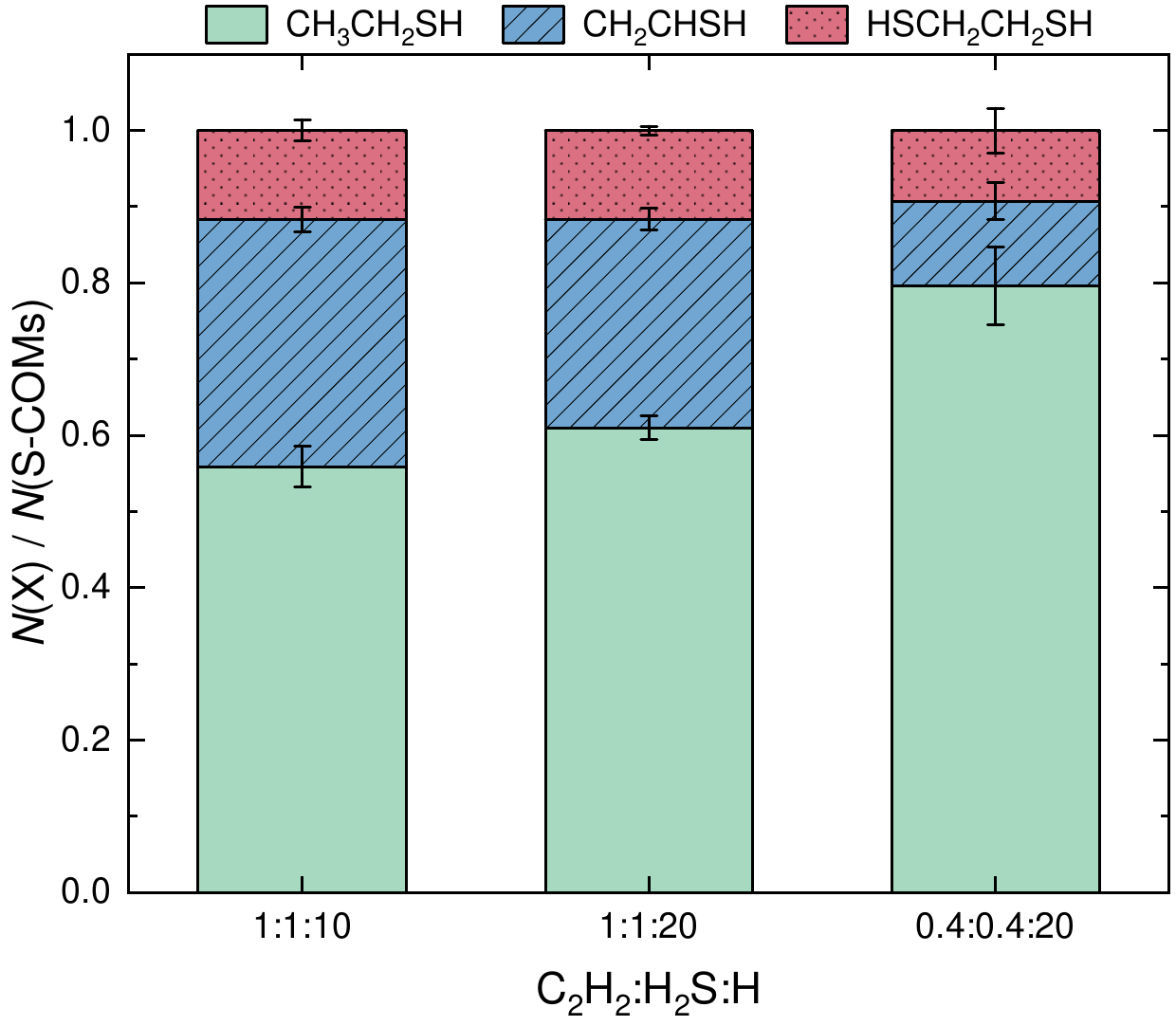}
\caption{Relative abundances of the sulfur-bearing COMs formed during deposition as a function of the \ce{C2H2}:\ce{H2S}:\ce{H} ratio. The abundances are shown with respect to the total yield of sulfurated COMs positively identified at the end of each experiment.}
\label{fig:NX_Nproducts}
\end{figure}

For the three main organic products, \ce{CH3CH2SH}, \ce{CH2CHSH}, and \ce{HSCH2CH2SH}, the dependence of the product yield with the molecule-to-hydrogen ratio is shown in Figure \ref{fig:NX_Nproducts}. The products' relative column densities are derived from the mass signals of their molecular ions using Equation \ref{eq:N_QMS} and the parameters in Table \ref{tab:params_QMS}. Thioacetaldehyde and thioketene (\ce{CH3CHS} and \ce{CH2CS}, respectively) are not included in this analysis since they are tentative detections and minor products. For all explored deposition conditions, the most abundant S-bearing organic molecule resulting from the interaction of \ce{C2H2} with \ce{H2S} and \ce{H} atoms is \ce{CH3CH2SH}. The predominance of this species is in line with the conclusions drawn by \citet{Shingledecker2022} from their calculations and the computational results from this work. Once \ce{C2H2} is hydrogenated to form \ce{C2H3}, it can react barrierlessly with \ce{SH} radicals to form \ce{CH2CHSH}, which can subsequently be barrierlessly hydrogenated to form \ce{CH3CH2SH}. The production of \ce{CH3CH2SH} can therefore proceed very efficiently as long as there are \ce{H} atoms in the vicinity available to react, and hence is increasingly favored for higher H-to-molecule ratios. Thus, our experiments show that \ce{CH3CH2SH} acts as a sink in the chemical network explored here, meaning that the sulfur budget in this network will be mostly locked away into ethanethiol at the expense of the other products---in agreement with the theoretical predictions \cite{Shingledecker2022}. These conclusions remain valid irrespective of the formation route behind \ce{CH2CHSH}, and thus are not exclusively dependent on the network being initiated by the interaction between \ce{C2H3} and \ce{SH}. \ce Indeed, as suggested by \citet{Shingledecker2022}, the absence of \ce{CH2CHSH} in interstellar sources despite dedicated attempts to identify it towards Sgr B2(N2) \cite{Martin-Drumel2019}, alongside the detections of \ce{CH3CH2SH} towards both Orion KL and the G+0.693-0.027 molecular cloud \cite{Koleniskova2014, RodriguezAlmeida2021}, might be related to the effective chemical conversion of the former to the latter. 

Thioketene (\ce{CH2CS}) was also detected in interstellar environments, towards the cold core TMC-1 \cite{Cernicharo2021}. The reaction routes proposed in this work could in principle contribute to forming it in the solid phase, albeit to a small extent. Nonetheless, its astronomical detection reinforces the relevance of investigating chemical networks that lead to sulfur-bearing organics with two carbon atoms. 
\section{Conclusions}
\label{sec:conc}

In the present work, we explore the solid-state chemistry resulting from the interaction of \ce{C2H2}, \ce{H2S}, and \ce{H} atoms in interstellar ices at 10 K by means of laboratory experiments combined with theoretical calculations. The investigated chemical network is summarized in Figure \ref{fig:scheme}, and our main findings are listed below:
\begin{itemize}
    \item The codeposition of \ce{C2H2}, \ce{H2S}, and \ce{H} leads to the formation of several sulfur-bearing species at 10 K via radical-induced reactions involving \ce{SH}. We securely identify the products \ce{CH3CH2SH}, \ce{CH2CHSH}, \ce{HSCH2CH2SH}, \ce{H2S2}, and tentatively \ce{CH3CHS} and \ce{CH2CS}, by using infrared spectroscopy and mass spectrometry.
    \item Calculations at the  M062X/def2-TZVP level of theory bechmarked with  CCSD(T)-F12/AUG-CC-PVTZ predict an activation barrier of $\sim$758 K for the radical-molecule reaction $\ce{C2H2} + \ce{SH} \to \ce{C2H2SH}$, which is deterrent under molecular clouds conditions. Thus, the chemical network is likely initiated by the interaction between \ce{C2H3} and \ce{SH} radicals.
    \item The product \ce{CH2CHSH} plays an important role as an intermediate as it can be hydrogenated to form \ce{CH3CH2SH} or potentially converted to its isomer \ce{CH3CHS}. Given enough H-atom availability, it will be largely consumed to produce more stable species.
    \item For all explored deposition conditions, the main product formed is \ce{CH3CH2SH}, with percentage yield with respect to the sum of S-bearing COMs ranging from $\sim$56\% to $\sim$80\%. The yield of \ce{CH3CH2SH} increases with the \ce{H} fraction due to its efficient formation through a series of barrierless hydrogenation reactions. It therefore acts as a sulfur sink in the present chemical network, being preferably formed at the expense of the other products. 
\end{itemize}

Astronomical detections of \ce{CH3CH2SH} and \ce{CH2CS} in the submillimiter range evince the importance of exploring the formation mechanisms of sulfur-bearing organic molecules with two carbon atoms under interstellar cloud conditions. In this work we experimentally investigate solid-phase formation routes that enrich the ice with the aforementioned products, in particular during the early cloud stages. Nonetheless, further laboratory, observational, and modelling works are warranted to better constrain the ice abundances of \ce{C2H2} and \ce{H2S} and the yield of products in more representative interstellar ices.

\begin{acknowledgement}

This work is dedicated to the memory of Prof. Dr. Harold Linnartz. The authors thank the Danish National Research Foundation through the Center of Excellence “InterCat” (Grant agreement no.: DNRF150) and the Netherlands Research School for Astronomy (NOVA) for their support. KJC acknowledges the support by the NWO via a VENI fellowship (VI.Veni.212.296). JER acknowledges the support by the Horizon Europe Framework Programme's (HORIZON) Marie Skłodowska-Curie, grant agreement No 101149067.
\end{acknowledgement}


\bibliography{mybibfile}

\end{document}